\documentclass[preprint,onecolumn,superscriptaddress]{revtex4}

\usepackage{graphicx}
\usepackage{dcolumn}
\usepackage{bm}
\usepackage{amssymb}
\usepackage{amsfonts}
\usepackage{latexsym}
\usepackage{enumerate}
\usepackage[dvipdfmx]{color} 
\usepackage{amsmath}
\usepackage[dvipdfmx]{epsfig}
\usepackage{times}
\usepackage{algorithm}
\usepackage{algorithmic}
\usepackage{cases}

\newcommand{\red}{\color{black}}
\newtheorem{theorem}{Theorem}
\newtheorem{lemma}{Lemma}

\begin{document}
\widetext
\title{Resource-efficient 
verification of quantum computing using Serfling's bound}
YITP-18-58
\author{Yuki Takeuchi}
\email{yuki.takeuchi.yt@hco.ntt.co.jp}
\affiliation{Graduate School of Engineering Science, Osaka University, Toyonaka, Osaka 560-8531, Japan}
\affiliation{NTT Communication Science Laboratories, NTT Corporation, 3-1 Morinosato Wakamiya, Atsugi-shi, Kanagawa 243-0198, Japan}
\author{Atul Mantri}
\email{amantri@exseed.ed.ac.uk}
\affiliation{Singapore University of Technology and Design, 8 Somapah Road, Singapore 487372}
\affiliation{Centre for Quantum Technologies, National University of Singapore, Block S15, 3 Science Drive 2, Singapore 117543}
\affiliation{\red School of Informatics, University of Edinburgh,10 Crichton Street, Edinburgh EH8 9AB, UK}
\author{Tomoyuki Morimae}
\email{tomoyuki.morimae@yukawa.kyoto-u.ac.jp}
\affiliation{Yukawa Institute for Theoretical Physics, Kyoto University, Kitashirakawa Oiwakecho, Sakyo-ku, Kyoto 606-8502, Japan}
\affiliation{JST, PRESTO, 4-1-8 Honcho, Kawaguchi, Saitama, 332-0012, Japan}
\author{Akihiro Mizutani}
\email{Mizutani.Akihiro@dy.MitsubishiElectric.co.jp}
\affiliation{Graduate School of Engineering Science, Osaka University, Toyonaka, Osaka 560-8531, Japan}
\affiliation{Mitsubishi Electric Corporation, Information Technology R\&D Center, 5-1-1 Ofuna, Kamakura-shi, Kanagawa, 247-8501, Japan}
\author{Joseph F. Fitzsimons}
\email{joe@horizonquantum.com}
\affiliation{Singapore University of Technology and Design, 8 Somapah Road, Singapore 487372}
\affiliation{Centre for Quantum Technologies, National University of Singapore, Block S15, 3 Science Drive 2, Singapore 117543}
\affiliation{\red Horizon Quantum Computing, 79 Ayer Rajah Crescent, \#03-01 BASH, Singapore 139955}

\begin{abstract}
\noindent Verifying quantum states is central to certifying the correct operation of various quantum information 
processing tasks. In particular, in measurement-based quantum computing,
checking whether correct graph states are generated is essential 
for reliable quantum computing. Several verification protocols for graph states have been proposed, but none of these are particularly resource efficient: multiple copies are
required to extract a single state that is guaranteed to be close to the ideal 
{\red one}. {\red The} best protocol currently known requires
$O(n^{15})$ copies of the state, where $n$ is the size of the graph state.
In this paper, we construct a significantly more 
resource-efficient verification protocol for graph states
that only requires $O(n^5\log{n})$ copies.
The key idea is to employ Serfling's bound, which is
a probability inequality in classical statistics. 
Utilizing Serfling's bound also enables us to generalize our protocol
for qudit and continuous-variable graph states.
Constructing a resource-efficient verification protocol
for {\red them} is non-trivial.
For example, the previous verification protocols for qubit graph states that use the
quantum de Finetti theorem cannot be
generalized to qudit and continuous-variable graph states without
tremendously increasing the resource overhead.
This is because the overhead caused by the quantum de Finetti theorem 
depends on the local dimension.
On the other hand, in our protocol, the resource overhead is independent of the local dimension, and therefore generalizing to qudit or continuous-variable graph states does not increase the overhead.
The flexibility of Serfling's bound also makes our protocol robust: 
our protocol accepts slightly noisy but still useful graph states.
\end{abstract}
\maketitle

\medskip
\noindent{\bf\large INTRODUCTION}\\
The verification of quantum states plays an important role in ensuring the integrity of a number of
quantum technologies including quantum computing, quantum cryptography, and quantum simulations. 
Graph states are a particularly important class of quantum states, since they can be resource states for measurement-based quantum computation (MBQC)~\cite{RB01,RBB03,ZZXS03,HEB04,MLGWRN06,GE07,RHG07,MM16}.
In the case of MBQC, the problem of verifying quantum computation reduces to simply certify the measurements and the resource state. In this paper, we consider the task of verifying these important resource states that include qubit graph states, qudit graph states, and continuous-variable (CV) graph states.

\begin{figure}[t]
\includegraphics[width=8cm, clip]{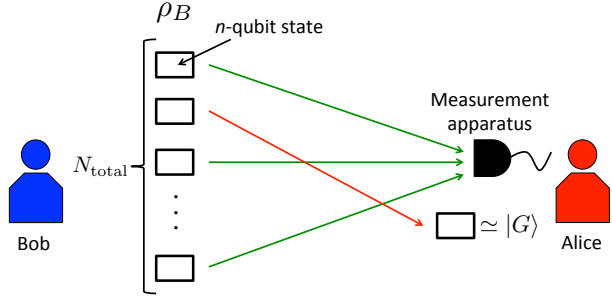}
\caption{The schematic shows the set-up considerd within this paper. 
Bob generates an $nN_{\rm total}$-qubit state $\rho_B$ that consists of $N_{\rm total}$ registers. Each register is represented by a box and contains $n$ qubits. Bob sends the state $\rho_B$ to Alice.
Alice randomly chooses $N_{\rm total}-1$ registers, and measures each qubit of them. If the measurement results satisfy certain conditions, the unmeasured single register is guaranteed to be close to the ideal graph state $|G\rangle$.
If Bob sends Alice each qubit of $\rho_B$ one by one,
and if Alice measures each qubit sequentially,
Alice does not need any quantum memory.
If we consider the qudit (qumode) case, the word ``qubit" in the above explanation should be replaced with ``qudit" (``qumode").}
\label{verification}
\end{figure}

Let us consider the following general setup (as shown in Fig.~\ref{verification}): 
Bob has a universal quantum computer and he can prepare arbitrary quantum states.
Alice, on the other hand, can only perform single-qubit measurements.
She does not have any quantum memory and the ability to apply entangling gate operations.
Therefore, Alice delegates the preparation of graph states to Bob.
Bob generates an $nN_{\rm total}$-qubit state 
$\rho_B$ and sends it to Alice.
The state $\rho_B$ consists of $N_{\rm total}$ registers, and each register
contains $n$ qubits. 
If Bob is honest, then the state of each register received by Alice is
the correct $n$-qubit graph state $|G\rangle$. 
In other words,
\begin{eqnarray}
\rho_B=|G\rangle\langle G|^{\otimes N_{\rm total}}.
\end{eqnarray}
If Bob is malicious, on the other hand,
$\rho_B$ can be any arbitrary $nN_{\rm total}$-qubit state.
Alice randomly chooses $N_{\rm total}-1$ registers and 
measures all of them.
In such a scenario, is it possible to construct a protocol such 
that if these measurement results
satisfy certain conditions,
then the state of the remaining (i.e., unmeasured) single register is 
guaranteed to be close to the ideal state $|G\rangle$?
If such a verification protocol is possible, Alice can safely
use the verified register for her desired 
MBQC. {\red If we consider the qudit (qumode) case, the word ``qubit" in the explanation should be replaced with ``qudit" (``qumode").}

This setup models usual experiments. Bob is {\red regarded to be} an experimental equipment
that is expected to generate many copies of
the $n$-qubit graph state $|G\rangle$,
and Alice is an experimentalist who has constructed the equipment.
She would like to check the correctness of the experimental
equipment.
In this case, it is unreasonable to assume that $\rho_B$ is 
any state, since the ``attack" by the experimental equipment is not
a malicious one, but is due to natural noise.
In other words,
\begin{eqnarray}
\label{rhoBI}
\rho_B=\mathcal{E}\left(|G\rangle\langle G|^{\otimes N_{\rm total}}\right),
\end{eqnarray}
where $\mathcal{E}$ is a completely positive trace-preserving (CPTP) map that represents
certain noise caused by the interaction between the experimental equipment and the environment.

Due to the inherent asymmetry between Alice and Bob, the verification setup shown in Fig.~\ref{verification} can also be considered as a
cloud quantum computing scenario~\cite{BFK09,MF13A}. 
{\red Imagine that Bob is} the owner of a company which provides a
quantum computing service over the cloud, and Alice as a (computationally weak) client who wants to
use the cloud service to perform her desired quantum operations. {\red One way} to achieve this task is {\red that Bob generates} graph states
and send them to Alice one qubit at a time. Alice, who can only perform single-qubit measurements,
measures each qubit, as it arrives, to perform
her MBQC. Since Alice does not trust Bob, she has to verify
the correctness of the graph states by herself.
This situation is well modeled by taking $\mathcal{E}$ in Eq.~(\ref{rhoBI}) to be a general CPTP map, since Bob is restricted to non-adaptive attacks, which is equivalent to maliciously preparing the initial state.

Several verification protocols, including ones based on self-testing, have been proposed for graph states~\cite{MTH17,TM17,FH17,SKM17,HM15,DA18,ZH18,M11}.
However, all the previous protocols encounter at least one of the three problems described below.
First, they are not resource efficient, i.e., $N_{\rm total}$ is large.
For example, protocols of Refs.~\cite{MTH17,TM17} use the idea of the quantum de Finetti theorem~\cite{LS15} to
make $\rho_B$ close to independent and identically distributed (i.i.d.) copies of a single fixed state, 
but this comes at the cost of large overhead.
Such protocols require $N_{\rm total}=O(n^{21})$ to complete the verification. 
There are some protocols that do not use the quantum de Finetti 
theorem~\cite{HM15,FH17,DA18,ZH18}, but nonetheless $N_{\rm total}=O(n^{15})$ (for details, see the ``Resource efficiency'' subsection in the RESULTS section). 

Second, although extensions to qudit or CV states are in principle possible for existing protocols~\cite{HM15,DA18,ZH18} (and in fact, extensions to qudit systems are explicitly mentioned in Refs.~\cite{HM15,DA18,ZH18}), extending the previous verification protocols~\cite{MTH17,TM17,SKM17,M11} to higher dimension without increasing $N_{\rm total}$ is highly non-trivial. In fact, in
the case when the quantum de Finetti theorem is used for a qudit system
with the qudit dimension $d\ge3$, for example, $N_{\rm total}$ depends on 
$\sqrt{\log{d}}$, and therefore if we take $d\to\infty$ to construct
a CV verification protocol, we come across the unpleasant consequence
that $N_{\rm total}\to\infty$.
Note that recently, in Ref.~\cite{ZH18}, the generalization to qudit systems has been made without increasing $N_{\rm total}=O(n^{15})$. However, this resource overhead is large. Resource-efficient verification protocols for qudit and CV states are indispensable.
In fact, there are many important qudit or CV
quantum states such as the photon orbital angular 
momentum states~\cite{DLBPA11} and the Gottesman-Kitaev-Preskill (GKP) 
states~\cite{GKP01}. Recently, large-scale CV entangled states composed of 
more than $10^4$ qumodes have been generated~\cite{YUASKSYYMF13}.

Finally, although many
previous protocols~\cite{HM15,DA18,SKM17,ZH18} have {\red an} error tolerance, it is low or its evaluation is insufficient (for details, see the ``Robustness" subsection in the RESULTS section). The error tolerance is an important feature of verification protocols {\red due to} the following reason: Even if Bob is honest,
the state $\rho_B$ that Alice receives might be slightly different from 
the ideal state $|G\rangle^{\otimes N_{\rm total}}$ because of 
the imperfections of Bob's operations and the channel noise between
Bob and Alice. 
These slightly perturbed states can still be useful
for fault-tolerant computation.
The acceptance of such states by Alice would lead to better verification protocols. A fault-tolerant verification 
protocol~\cite{FH17} that accepts noisy but still error-correctable states has 
been proposed, but it can be only used for the bipartite graph states.

In this paper, we introduce a new verification protocol
that overcomes all the three aforementioned problems.
First, we show that our protocol is significantly more resource efficient than previous approaches, i.e., $N_{\rm total}=O(n^5\log{n})$. (Recall that the best known previous protocols need $N_{\rm total}=O(n^{15})$~\cite{HM15,FH17,DA18,ZH18}.)
The key idea to achieve this efficiency
is by employing Serfling's bound~\cite{S74,TL17}. 
Serfling's bound is a probability inequality for the sum
in sampling without replacement, and is often used in classical statistics (for details,
see Lemma~\ref{randomsampling}). Serfling's bound has also been used in the security proofs of quantum key distribution~\cite{TL17,FFBLSTW12,LPTRG13,PRMTW18},
but to the best of our knowledge, {\red so far,} Serfling's bound has {\red never} been applied to the 
verification of quantum computation.

Second, our protocol can be generalized to qudit and CV graph states,
while maintaining resource efficiency. As we have explained, constructing a resource-efficient qudit or CV verification protocol is non-trivial.
The reason why we succeed in the construction is again because of
Serfling's bound; When we use Serfling's bound for a qudit system
with $d\ge3$, 
$N_{\rm total}$ is {\it independent} of $d$. 
Therefore, we can increase $d$ without increasing $N_{\rm total}$.
Note that our verification protocol for CV graph states can also be generalized to CV weighted hypergraph states. Hypergraph states are generalizations of graph states, and ``weighted" implies that a real number is associated with each hyperedge. The precise definition will be given in the ``CV weighted hypergraph states" subsection in the RESULTS section.
Several important CV states are weighted hypergraph states, such as the CV cluster state~\cite{MLGWRN06} and the CV weighted toroidal lattice state~\cite{MFP08}, and our protocol is useful for verifying these states.
We also point out that our verification protocol for CV graph states can be used to construct a verifiable blind quantum computing protocol with CV states. To the best of our knowledge, no verifiable blind CV quantum computing protocol was known previously except for the protocol of Ref.~\cite{LDTAF18}. The protocol of Ref.~\cite{LDTAF18} assumes that the malicious server is restricted to preparing i.i.d. copies of single-qumode states while the malicious server in our protocol can perform any CPTP map as an attack.

Finally, our protocol is robust to some extent against slight perturbations of quantum states.
For example, in previous protocols~\cite{HM15,DA18,SKM17,ZH18}, all $N_{\rm total}-1$ registers have to pass a test in order for Alice to accept the remaining state as correct, which means that slightly perturbed but still useful states are rejected by Alice.
In our protocol, on the other hand, Alice accepts even if some of the $N_{\rm total}-1$ registers fail a test. This relaxed acceptance criteria allows Alice to accept noisy but still useful resource states. As non-fault-tolerant small-scale quantum processors are becoming available~\cite{IBM1,IBM2,IBM3}, our protocols may be useful to verify these near-term quantum computers.

In addition to the verification protocols for graph states mentioned above, there are other approaches to verifiable quantum computing. The protocols in Refs.~\cite{FK17,ABEM17} use trap based techniques to perform verifiable blind quantum computing. In their protocols, a client is required to prepare single-qubit states whereas single-qubit measurements are required in our setup. In order to make the client classical, several multi-server protocols have been proposed~\cite{RUV13,M16,NV17,CGJV17,FHM18}. Particularly, Coladangelo {\it et al.} have recently constructed two resource-efficient protocols for a classical client to verifiably delegate quantum computing to two non-communicating but entangled quantum servers~\cite{CGJV17}.
Other multi-server protocols have also been proposed to make verification protocols device independent~\cite{HPF15,HH18}.
However, the assumption that servers do not communicate with each other is hard to impose in practice due to latency in real-world networks and the finite speed of quantum operations. Recently, Mahadev has shown that quantum computation can be verified by an entirely classical client even when only one quantum processor is available, under computational assumptions~\cite{M18}. Such protocols, however, necessitate extremely large quantum processors due to the relatively large key sizes necessary for cryptographic security.\\

\medskip
\noindent{\bf\large RESULTS}\\
This section is organized as follows: first, as preliminaries, we review the definitions of qudit graph states and their stabilizer operators. Second, we construct a stabilizer test as a sub-protocol of our verification protocol. With respect to CV weighted hypergraph states, we also review their definitions and construct a stabilizer test. After that, based on the stabilizer tests, we propose our verification protocol, which is the main result of this paper. We also discuss the resource efficiency and the error tolerance of our verification protocol. Finally, we generalize our verification protocol so that it can be used to verify multiple quantum states simultaneously.

\medskip
\noindent {\bf Qudit graph states}\\
A graph $G\equiv(V,E)$ is a pair of a set $V\equiv\{v_i\}_{i=1}^n$ of vertices and a set $E\equiv\{e_i\}_{i=1}^{|E|}$ of edges with $n\equiv|V|$. Here, $|V|$ and $|E|$ denote the number of elements of $V$ and $E$, respectively. Let $\{|k\rangle\}_{k=0}^{d-1}$ be an orthonormal basis in a $d$-dimensional Hilbert space, where $d(\ge2)$ is finite. A qudit graph state $|G_d\rangle$ that corresponds to $G$ is defined by
\begin{eqnarray}
|G_d\rangle\equiv\left(\prod_{(i,j)\in E}{CZ}_{ij}\right)|+_d\rangle^{\otimes n},
\end{eqnarray}
where
\begin{eqnarray}
|+_d\rangle\equiv\cfrac{1}{\sqrt{d}}\sum_{k=0}^{d-1}|k\rangle
\end{eqnarray}
is the $+1$ eigenvector of the generalized Pauli-$X$ operator
\begin{eqnarray}
X\equiv\sum_{k=0}^{d-1}|k+1\ ({\rm mod}\ d)\rangle\langle k|,
\end{eqnarray}
and 
\begin{eqnarray}
{CZ}_{ij}\equiv\sum_{k=0}^{d-1}\sum_{k'=0}^{d-1}{\rm exp}\left(i2\pi\cfrac{kk'}{d}\right)|kk'\rangle_{ij}\langle kk'|_{ij}
\end{eqnarray}
is a qudit analogue of the controlled-$Z$ $(CZ)$ gate acting on the $i$th and the $j$th qudits. It is easy to confirm that when $d=2$, a qudit graph state becomes a conventional qubit graph state. The $i$th stabilizer $g_i^{(d)}$ (with $1\le i\le n$) of $|G_d\rangle$ is given by
\begin{eqnarray}
g_i^{(d)}&\equiv&\left(\prod_{(i,j)\in E}{CZ}_{ij}\right)X_i\left(\prod_{(i,j)\in E}{CZ}_{ij}^\dag\right)\\
&=&X_i\prod_{v_j\in N^{(i)}}Z_j,
\end{eqnarray}
where $N^{(i)}$ is the set of neighbors of the $i$th vertex, and
\begin{eqnarray}
Z_j\equiv\sum_{k=0}^{d-1}e^{i2\pi k/d}|k\rangle\langle k|
\end{eqnarray}
is the generalized Pauli-$Z$ operator acting on the $j$th qudit. It is easy to check that $g_i^{(d)}|G_d\rangle=|G_d\rangle$.

\medskip
\noindent {\bf Stabilizer test for qudit graph states}\\
Consider a stabilizer test which is an essential sub-protocol of our verification protocol for qudit graph states.
Let $\rho$ be an $n$-qudit quantum state. We define the stabilizer test for $g_i^{(d)}$ on $\rho$ as follows: Alice measures $X_i$ and $Z_j$ for all $j\in N^{(i)}$. Let $x_i\in\{0,\ldots,d-1\}$ and $z_j\in\{0,\ldots,d-1\}$ be the measurement outcomes of $X_i$ and $Z_j$, respectively. We say that Alice passes the stabilizer test for $g_i^{(d)}$ on $\rho$ if 
\begin{eqnarray}
\label{passd}
x_i+\sum_{j\in N^{(i)}}z_j\equiv 0\ ({\rm mod}\ d).
\end{eqnarray}
Since the correct qudit graph state $|G_d\rangle$ always satisfies Eq.~(\ref{passd}), it passes the stabilizer test for $g_i^{(d)}$ with unit probability for all $i$.

\medskip
\noindent {\bf CV weighted hypergraph states}\\
A weighted hypergraph $G\equiv(V,E,\Omega)$ is a triple of a set $V\equiv\{v_i\}_{i=1}^n$ of vertices, a set $E\equiv\{e_i\}_{i=1}^{|E|}$ of hyperedges, and a set $\Omega\equiv\{\Omega_i\}_{i=1}^{|E|}$ of weights, where $n\equiv|V|$. Here, a hyperedge is a set of vertices, and $\Omega_i\in\mathbb{R}$ represents the weight of the $i$th hyperedge. An edge is a special case of the hyperedge when $|e|=2$, where $|e|$ denotes the number of vertices linked to the hyperedge $e$.
Let
\begin{eqnarray}
\label{G_CV}
|G_{\rm CV}\rangle\equiv\left(\prod_{j=1}^{|E|}{CZ}_{e_j}(\Omega_j)\right)|0_p\rangle^{\otimes n}
\end{eqnarray}
be a CV weighted hypergraph state corresponding to $G$, where $|0_p\rangle_{v_i}$ is a phase-squeezed state corresponding to the $i$th vertex, i.e., the eigenvector of the phase quadrature operator $\hat{p}\equiv-i(\hat{a}-\hat{a}^\dag)/\sqrt{2}$ corresponding to the eigenvalue $0$, $\hat{a}$ $(\hat{a}^\dag)$ is the boson annihilation (creation) operator,
\begin{eqnarray}
\label{CZhyper}
{CZ}_{e_j}(\Omega_j)\equiv{\rm exp}\left(i\Omega_j\prod_{v_i\in e_j}\hat{x}_i\right)
\end{eqnarray}
is a CV analogue of the $CZ$ gate acting on qumodes corresponding to vertices in the hyperedge $e_j$, and $\hat{x}_i\equiv(\hat{a}_i+\hat{a}_i^\dag)/\sqrt{2}$ is the amplitude quadrature operator acting on the $i$th qumode. From Eqs.~(\ref{G_CV}) and (\ref{CZhyper}), it can be seen that when $|e_j|=2$ for all $j$, a CV weighted hypergraph state becomes a CV weighted graph state. Since a hypergraph has at most $2^n-1$ hyperedges, the time required to generate a hypergraph state is at most $O(2^n)$.
Hereafter, we assume that $|E|={\rm poly}(n)$, because states with superpolynomial scaling in $n$ in general require greater than polynomial time to generate, which is considered inefficient.
The $i$th stabilizer $g_i^{({\rm CV})}$ (with $1\le i\le n$) of $|G_{\rm CV}\rangle$ is defined by
\begin{eqnarray}
\label{stabilizerCV}
g_i^{({\rm CV})}\equiv\left(\prod_{j=1}^{|E|}{CZ}_{e_j}(\Omega_j)\right)\hat{p}_i\left(\prod_{j=1}^{|E|}{CZ}_{e_j}(-\Omega_j)\right),
\end{eqnarray}
where $\hat{p}_i$ is the phase quadrature operator acting on the $i$th qumode.
Let $E^{(i)}$ be the set of hyperedges that contain the $i$th vertex $v_i$. From Eq.~(\ref{stabilizerCV}),
\begin{eqnarray}
g_i^{({\rm CV})}&=&\hat{p}_i-\sum_{e_j\in E^{(i)}}\Omega_j\prod_{v_k\in e_j-\{v_i\}}\hat{x}_k,
\end{eqnarray}
where we have used the Baker-Hausdorff lemma and the commutation relation $[\hat{x},\hat{p}]=i$. It is easy to check that $g_i^{({\rm CV})}|G_{\rm CV}\rangle=0|G_{\rm CV}\rangle$. Since the quadrature operators can be measured using homodyne measurements, and $\Omega$ are known, the measurement of $g_i^{({\rm CV})}$ can be accomplished using only homodyne measurements, which are single-qumode measurements.

\medskip
\noindent {\bf Stabilizer test for CV weighted hypergraph states}\\
The stabilizer test for CV weighted hypergraph states forms an essential sub-protocol of our CV verification protocol.
Let $\rho$ be an $n$-qumode quantum state. We define the stabilizer test for $g_i^{({\rm CV})}$ on $\rho$ as follows: Alice measures $\hat{p}_i$ on the $i$th qumode and $\hat{x}_k$ on all qumodes in $\cup_{e_j\in E^{(i)}}e_j-\{v_i\}$.
Let $p_i\in\mathbb{R}$ and $x_k\in\mathbb{R}$ be the measurement outcomes of $\hat{p}_i$ and $\hat{x}_k$, respectively. We say that Alice passes the stabilizer test for $g_i^{({\rm CV})}$ on $\rho$ if 
\begin{eqnarray}
\label{passc}
p_i-\sum_{e_j\in E^{(i)}}\Omega_j\prod_{v_k\in e_j-\{v_i\}}x_k=0.
\end{eqnarray}
Here, we assume that measurements are infinitely accurate. However, as shown in the ``Robustness'' subsection in the RESULTS section, we can relax this assumption to some extent.
Since we assume that $|E|={\rm poly}(n)$, Alice can calculate the left-hand side of Eq.~(\ref{passc}) in classical polynomial time.
The correct CV weighted hypergraph state $|G_{\rm CV}\rangle$ passes the stabilizer test for $g_i^{({\rm CV})}$ with unit probability for all $i$ because the correct CV weighted hypergraph state $|G_{\rm CV}\rangle$ always satisfies Eq.~(\ref{passc}).

\begin{figure}[H]
\includegraphics[width=18cm, clip]{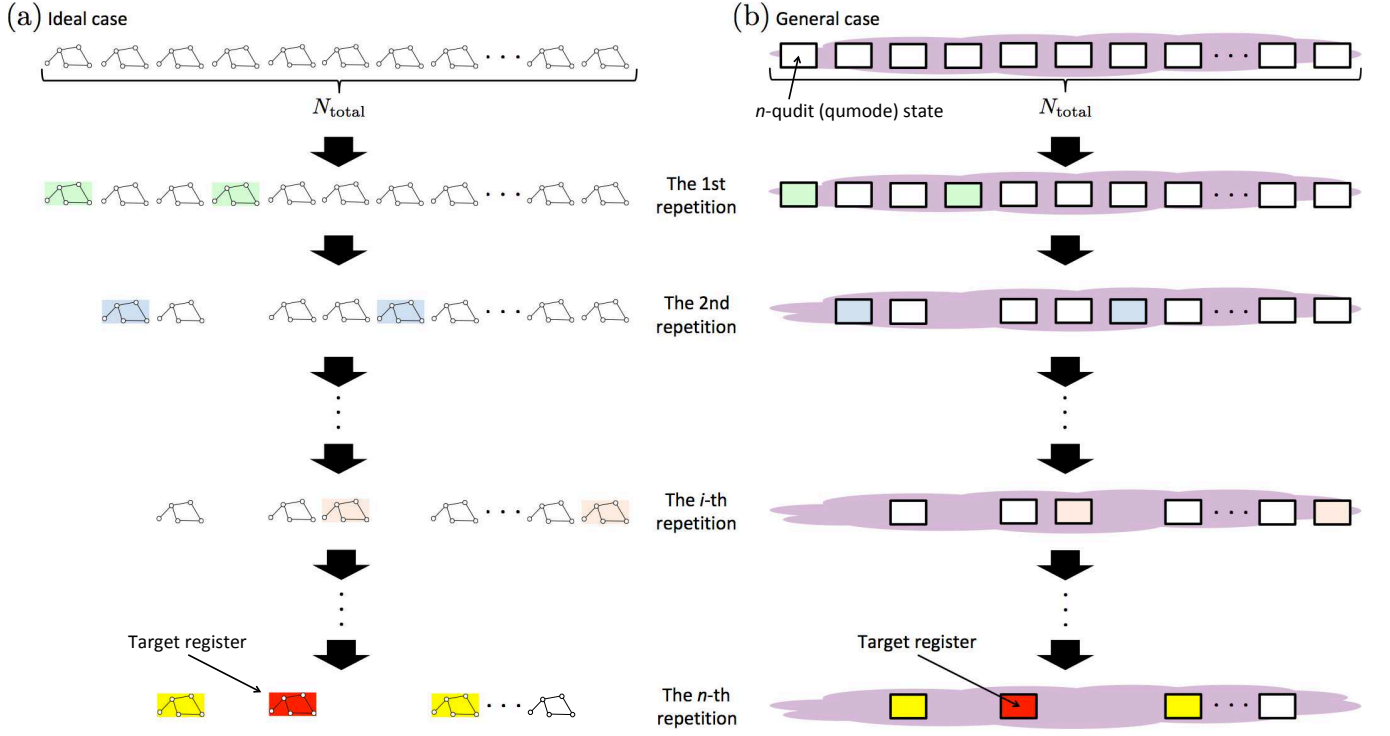}
\caption{The schematic shows our verification protocol. (a) The ideal case, i.e., $\rho_B=|G\rangle\langle G|^{\otimes N_{\rm total}}$, where $|G\rangle\in\{|G_d\rangle,|G_{\rm CV}\rangle\}$. In this figure, $|G\rangle$ is represented by the five-vertex graph $(n=5)$. In the first repetition of step 2 of our protocol, Alice chooses $N_{\rm test}$ registers uniformly and independently at random from $N_{\rm total}$ registers. The chosen $N_{\rm test}$ registers are highlighted in green color. Alice performs the stabilizer test for $g_1\in\{g_1^{(d)},g_1^{({\rm CV})}\}$ on each of these chosen green registers. In the second repetition of step 2 of our protocol, Alice chooses $N_{\rm test}$ registers uniformly and independently at random from the remaining $N_{\rm total}-N_{\rm test}$ registers. These chosen $N_{\rm test}$ registers are highlighted in blue color. Alice performs the stabilizer test for $g_2\in\{g_2^{(d)},g_2^{({\rm CV})}\}$ on each of these chosen blue registers. Alice repeats the same procedure $n$ times.
In the $n$-th repetition, i.e., in the last repetition, Alice chooses $N_{\rm test}$ registers uniformly and independently at random from the remaining $N_{\rm total}-(n-1)N_{\rm test}$ registers. These chosen $N_{\rm test}$ registers are highlighted in yellow color. Alice performs the stabilizer test for $g_n\in\{g_n^{(d)},g_n^{({\rm CV})}\}$ on each of these chosen yellow registers.
Finally, Alice chooses one register, which we call the target register, from the remaining $N_{\rm total}-nN_{\rm test}$ registers. The target register is highlighted in red color. If the results of the stabilizer tests satisfy a certain condition (see step 4 of our protocol), the state $\rho_{\rm tgt}$ of the target register is guaranteed to be close to $|G\rangle$ with high probability.
(b) In general, $\rho_B$ can be any $nN_{\rm total}$-qudit (qumode) state. The state $\rho_B$ consists of $N_{\rm total}$ registers, and each register contains $n$ qudits (qumodes). Each register is represented by a box. Any entanglement can be generated amongst registers, which is indicated by the purple ``cloud" behind boxes.}
\label{soundfig}
\end{figure}

\medskip
\noindent {\bf Main protocol}\\
Our verification protocol for qudit graph states (CV weighted hypergraph states) runs as follows (see Fig.~\ref{soundfig}):
\begin{enumerate}
\item Bob sends an $nN_{\rm total}$-qudit (qumode) state $\rho_B$ to Alice as shown in Fig.~\ref{verification}. The state $\rho_B$ consists of $N_{\rm total}$ registers, and each register stores $n$ qudits (qumodes). If Bob is honest, the state of each register is $|G_d\rangle$ $(|G_{\rm CV}\rangle)$, i.e., $\rho_B=|G_d\rangle\langle G_d|^{\otimes N_{\rm total}}$ ($\rho_B=|G_{\rm CV}\rangle\langle G_{\rm CV}|^{\otimes N_{\rm total}}$). On the other hand, if he is malicious, $\rho_B$ can be any arbitrary state.

\item Alice repeats the following for $i=1,\ldots,n$:
she chooses $N_{\rm test}$ registers from the remaining $N_{\rm total}-(i-1)N_{\rm test}$ registers independently and uniformly at random, and then she performs the stabilizer tests for $g_i^{(d)}$ $(g_i^{({\rm CV})})$ on each of them. Let $N_{{\rm pass},i}$ be the number of registers that pass the stabilizer test for $g_i^{(d)}$ $(g_i^{({\rm CV})})$.

\item Alice uniformly and randomly chooses a single register from the remaining $N_{\rm total}-nN_{\rm test}$ registers that were not used for the stabilizer tests in step 2. We call the chosen single register the target register. Therefore, the averaged state of the target register is $\rho_{\rm tgt}\equiv\sum_{i}\rho_i/(N_{\rm total}-nN_{\rm test})$, where $\rho_i$ is the $i$th remaining register. All the other $N_{\rm total}-nN_{\rm test}-1$ registers are discarded.

\item If
\begin{eqnarray}
\label{threshold}
\sum_{i=1}^nN_{{\rm pass},i}\ge \left(n-\cfrac{1}{2n}\right)N_{\rm test},
\end{eqnarray}
she uses the target register for her MBQC, otherwise she discards the target register.
\end{enumerate}
We will later show that our verification protocol gives the lower bound on the fidelity between the state $\rho_{\rm tgt}$ of the target register and the ideal graph state $|G_d\rangle$ $(|G_{\rm CV}\rangle)$ (see Theorem~\ref{soundness}).

Note that in the above protocol, no quantum memory is needed for Alice. This is because Bob sends each qubit of $\rho_B$ one by one to Alice, and she randomly chooses her action from the stabilizer tests, MBQC on the target register, and discarding. Most importantly, our protocol does not assume any i.i.d. property on the quantum state $\rho_B$. In other words, $\rho_B$ can be any state, and we do not assume that $\rho_B=\sigma^{\otimes N_{\rm total}}$, where $\sigma$ is an $n$-qudit (qumode) state.

To show that our protocol is valid, we now show its completeness and soundness. Intuitively, if Alice accepts the correct quantum state with high probability, we say that the verification protocol has the completeness. In other words, the completeness means that Alice does not mistakenly reject the correct quantum state. On the other hand, if the verification protocol guarantees that the accepted quantum state is close to the correct state with high probability, we say that the verification protocol has the soundness. That is, the soundness means that Alice does not mistakenly accept any quantum state that is far from the ideal state.
It is not difficult to show the completeness of our protocol. In fact, 
when Bob is honest, i.e., when he sends $|G_d\rangle^{\otimes N_{\rm total}}$ $(|G_{\rm CV}\rangle^{\otimes N_{\rm total}})$ to Alice, she uses the target register for her MBQC in step 4 with unit probability, because $\sum_{i=1}^nN_{{\rm pass},i}=nN_{\rm test}$. With respect to the soundness, the following theorem holds:
\begin{theorem}[Soundness]
\label{soundness}
Let $N_{\rm pass}\equiv\sum_{i=1}^nN_{{\rm pass},i}$.
If we set $N_{\rm total}=2nN_{\rm test}$ and $N_{\rm test}=\lceil5n^4\log{n}/32\rceil$, the $n$-qudit (qumode) averaged state $\rho_{\rm tgt}$ of the target register (over all random selections) satisfies, with probability at least $1-n^{1-5c/64}$,
\begin{eqnarray}
\langle G|\rho_{\rm tgt}|G\rangle\ge1-\cfrac{2\sqrt{c}}{n}-2n\left(1-\cfrac{N_{\rm pass}}{n\lceil5n^4\log{n}/32\rceil}\right),
\end{eqnarray}
where $\lceil\cdot\rceil$ is the ceiling function, $c$ is any constant satisfying $64/5<c<(n-1)^2/4$, $n\ge9$, and $|G\rangle\in\{|G_d\rangle,|G_{\rm CV}\rangle\}$.
\end{theorem}
We defer the detailed proof of Theorem~\ref{soundness} to the Supplementary Information Section A.
A brief explanation of the proof is given in the METHODS section.

If Eq.~(\ref{threshold}) holds, Theorem~\ref{soundness} gives the non-trivial lower bound:
\begin{eqnarray}
\label{nontrivial}
\langle G|\rho_{\rm tgt}|G\rangle\ge1-\cfrac{2\sqrt{c}+1}{n}.
\end{eqnarray}
Hence Theorem~\ref{soundness} shows the soundness of our verification protocol.

\medskip
\noindent {\bf Resource efficiency}\\
To show the resource efficiency of our protocol, we compare it with the verification protocol of Ref.~\cite{DA18}, which can verify any graph state and is one of the most efficient protocols currently known.
For simplicity, let us consider the situation where the quantum state $\rho_B$ generated by Bob consists of $N_{\rm total}-1$ ideal states $|G\rangle^{\otimes (N_{\rm total}-1)}$ and a single incorrect $n$-qudit (qumode) state $\eta$. In other words, $\rho_B=P[(|G\rangle\langle G|)^{\otimes (N_{\rm total}-1)}\otimes\eta]$, where $P$ is a permutation operator for registers. Bob knows that how $P$ permutes $(|G\rangle\langle G|)^{\otimes (N_{\rm total}-1)}\otimes\eta$ while Alice does not. In this case, since $N_{\rm pass}\ge nN_{\rm test}-1$ holds with unit probability, Eq.~(\ref{nontrivial}) holds with probability at least $1-n^{1-5c/64}$.
In order to satisfy this statement, our verification protocol requires $N_{\rm total}=2n\lceil 5n^4\log{n}/32\rceil$ registers. The verification protocol of Ref.~\cite{DA18} guarantees the fidelity $1-1/(\alpha M)$ with probability at least $1-\alpha$ using $(N_{\rm total}=)M$ registers. Here, $\alpha$ is an upper bound on the probability of a quantum state whose fidelity is less than $1-1/(\alpha M)$ being selected as the target register. Therefore, if we require the protocol of Ref.~\cite{DA18} to achieve the same fidelity and the probability as ours,
\begin{eqnarray}
\begin{cases}
1-\cfrac{1}{\alpha M}=1-\cfrac{2\sqrt{c}+1}{n} & \\
\alpha=n^{1-5c/64} &
\end{cases}
\end{eqnarray}
have to be satisfied. From these equations, we obtain 
\begin{eqnarray}
M=\cfrac{n^{5c/64}}{2\sqrt{c}+1}.
\end{eqnarray}
For example, if $c=192$, $M=O(n^{15})$, which should be compared with our resource overhead $N_{\rm total}=O(n^5\log{n})$.
In general, if $c>64$, the order of $M$ is larger than that of $N_{\rm total}$, because if $c>64$, $M=O(n^t)$ and
\begin{eqnarray}
t=\cfrac{5c}{64}>\cfrac{5\times64}{64}=5
\label{99}
\end{eqnarray}
while $N_{\rm total}=O(n^5\log{n})$.
(Note that $n\ge 18$ is required to satisfy $c>64$ because $c<(n-1)^2/4$.)
A similar argument holds for other previous protocols~\cite{HM15,FH17,ZH18}.

The reason why our protocol is more efficient than that of Ref.~\cite{DA18} can be explained as follows: the fidelity $F'=1-1/(\alpha M)$ of their protocol depends on $M$ and $\alpha$, and therefore, $M$ has to be increased in order to decrease the order of $\alpha$ while keeping the order of $1-F'$ same. Our fidelity, $F=1-(2\sqrt{c}+1)/n$, also depends on $N_{\rm total}$ and the probability, because $n$ does, but our fidelity also contains the constant $c$, which is independent of $N_{\rm total}$.
Therefore, by increasing $c$ instead of increasing $N_{\rm total}$, we can improve the order of the probability without increasing the order of $1-F$.
The existence of such an extra parameter $c$ is the advantage of using Serfling's bound. (The constant $c$ comes from $\nu$ of Serfling's bound via $\nu=\sqrt{c}/n^2$.)
Note that in general, it is difficult to compare our protocol with previous protocols without performing experiments. This is because the parameter $N_{\rm pass}$ in Theorem~\ref{soundness} is determined by experiment.
In addition, it is also difficult to compare our protocol with other approaches mentioned in the last paragraph in the INTRODUCTION section. This is because their definitions of the soundness (verifiability) is different from ours. Our definition of the soundness relies on the fidelity between the ideal resource state and the actual one. On the other hand, their definitions do not (directly) rely on the fidelity.

\medskip
\noindent {\bf Robustness}\\
To investigate the error tolerance of our protocol, consider a simple example where all registers of $\rho_B$ are in the same state $\sigma$, and the state $\sigma$ is a slightly deviated state from the ideal state $|G\rangle$:
\begin{eqnarray}
\sigma=\left(1-\epsilon\right)|G\rangle\langle G|+\epsilon \eta, 
\end{eqnarray}
where $0<\epsilon<1$, and
$\eta$ is any $n$-qudit (qumode) state. In other words,
\begin{eqnarray}
\rho_B=\sigma^{\otimes N_{\rm total}}. 
\label{deviate}
\end{eqnarray}
When $\epsilon=1/{\rm poly}(n)$, such $\sigma$ is still useful for quantum computing, because
\begin{eqnarray}
\Big|{\rm Tr}[A\sigma]-{\rm Tr}[A|G\rangle\langle G|]\Big|&\le&\Big|\Big|\left(1-\epsilon\right)|G\rangle\langle G|+\epsilon \eta-|G\rangle\langle G|\Big|\Big|\\
&\le&2\sqrt{\epsilon},
\end{eqnarray}
where $||\cdot||$ is the trace norm, for any positive operator-valued measure (POVM) element $A$. Therefore, the output probability distribution of MBQC on $\sigma$ is close to that on $|G\rangle$ within the $1/{\rm poly}(n)$ error, which is sufficient enough to solve, for example, bounded error quantum polynomial time (BQP) decision problems.

When $\epsilon<O(1/(n^5\log{n}))$, our protocol can accept $\rho_B$ of Eq.~(\ref{deviate}) with almost unit probability.
In fact, by a direct calculation, our protocol accepts $\rho_B$ with the probability
\begin{eqnarray}
\label{success}
p_{\rm acc}\equiv\sum_{k=0}^{\lfloor N_{\rm test}/(2n)\rfloor}\binom{nN_{\rm test}}{k}(1-\epsilon)^{nN_{\rm test}-k}\epsilon^k,
\end{eqnarray}
where $\lfloor\cdot\rfloor$ is the floor function. From $p_{\rm acc}>(1-\epsilon)^{nN_{\rm test}}$ and $nN_{\rm test}=O(n^5\log{n})$, if $\epsilon<O(1/(n^5\log{n}))$, $p_{\rm acc}$ approaches $1$ in the limit of large $n$.

The acceptance probability $p_{\rm acc}$ of our protocol is higher than that of Ref.~\cite{DA18}, suggesting that our protocol is more robust than that of Ref.~\cite{DA18}.
In fact, the protocol of Ref.~\cite{DA18} accepts $\rho_B$ of Eq.~(\ref{deviate}) with the probability
\begin{eqnarray}
p'_{\rm acc}\equiv(1-\epsilon)^{M-1}.
\end{eqnarray}
Therefore, if $c>64$,
\begin{eqnarray}
p_{\rm acc}>(1-\epsilon)^{nN_{\rm test}}\ge(1-\epsilon)^{M-1}=p'_{\rm acc},
\end{eqnarray}
where we have used Eq.~(\ref{success}) and the fact that 
$nN_{\rm test}\le M-1$, which is asymptotically true for large $n$ when $c>64$, because 
$nN_{\rm test}=N_{\rm total}/2=O(n^5\log{n})$
and $M=O(n^t)$ where $t>5$ from Eq.~(\ref{99}).

For simplicity,
we have considered the tensor product of
the same state $\sigma$ as shown in Eq.~(\ref{deviate}).
However,  
it is easy to confirm that
a similar argument holds even when small entanglement is created among registers of $\rho_B$. Furthermore, since the errors considered above can also be treated as errors in measurements, we can relax the assumption where measurements are ideal to some extent.

\medskip
\noindent {\bf Verification of multiple quantum states}\\
In step 3 of our protocol, Alice chooses a single register, which we call the target register. What happens if she chooses $\tilde{n}$ registers, instead of a single register? We can show the following theorem:
\begin{theorem}
\label{theorem2}
Let $N_{\rm pass}\equiv\sum_{i=1}^nN_{{\rm pass},i}$.
If we set $N_{\rm total}=2nN_{\rm test}$ and $N_{\rm test}=\lceil5n^4\log{n}/32\rceil$, the averaged state $\tilde{\rho}_{\rm tgt}$ of $\tilde{n}$ target registers (over all random selections) satisfies, with probability at least $1-n^{1-5c/64}$,
\begin{eqnarray}
\nonumber
&&\langle G|^{\otimes\tilde{n}}\tilde{\rho}_{\rm tgt}|G\rangle^{\otimes\tilde{n}}\\
&\ge&1-\cfrac{(2\sqrt{c}+2n^2-2nN_{\rm pass}/{\red N_{\rm test}})\tilde{n}{\red N_{\rm test}}}{\red nN_{\rm test}-(\tilde{n}-1)},
\end{eqnarray}
where $\lceil\cdot\rceil$ is the ceiling function, $c$ is any constant satisfying $64/5<c<[n/\tilde{n}-32(\tilde{n}-1)/(5\tilde{n}n^4\log{n})-1]^2/4$, $n\ge9\tilde{n}$, $\tilde{n}=O(n^t)$, $t<1$, and $|G\rangle\in\{|G_d\rangle,|G_{\rm CV}\rangle\}$.
\end{theorem}
If Eq.~(\ref{threshold}) holds, Theorem~\ref{theorem2} gives the non-trivial lower bound:
\begin{eqnarray}
\label{nontrivial2}
\langle G|^{\otimes\tilde{n}}\tilde{\rho}_{\rm tgt}|G\rangle^{\otimes\tilde{n}}\ge1-\cfrac{(2\sqrt{c}+1)5\tilde{n}n^4\log{n}}{5n^5\log{n}-32(\tilde{n}-1)}.
\end{eqnarray}
Note that setting $\tilde{n}=1$ in Theorem~\ref{theorem2} results in Theorem~\ref{soundness}. This implies that Theorem~\ref{soundness} is a special case of Theorem~\ref{theorem2}. In several tasks, such as sampling problems, Alice would like to have several copies of graph states. Theorem~\ref{theorem2} is useful in such situations.
In simple terms, Eq.~(\ref{nontrivial2}) implies that Alice can obtain $\tilde{n}=O(n^t)$ quantum states with the fidelity $1-O(1/n^{1-t})$ using $N_{\rm total}=O(n^5\log{n})$ registers.
For verifying a single quantum state, the resource overhead $O(n^{5-t}\log{n})$ of our protocol is almost the same as that $O(n^4\log{n})$ of some previous device-independent multi-server verification protocols~\cite{HPF15,HH18}. Particularly, in the protocol of Ref.~\cite{HPF15}, the Azuma-Hoeffding bound~\cite{H63,A67} is used to achieve such overhead while we use Serfling's bound.
A proof of Theorem~\ref{theorem2} is provided in the Supplementary Information Section B.\\

\medskip
\noindent{\bf\large DISCUSSION}\\
We have proposed an efficient and robust verification protocol for any qudit graph state and any polynomial-time-generated CV weighted hypergraph state using Serfling's bound. Our protocol is much more efficient than the existing verification protocols when the size of the quantum state, the guaranteed fidelity, and the probability are sufficiently large.

Our analysis which is based on Serfling's bound is not directly applicable to the verification protocols for qubit hypergraph states~\cite{MTH17,TM17,ZH18}. This is mainly due to two reasons: a) in our analysis, the required number $N_{\rm total}$ of registers is proportional to the number of measurement settings, which is $n$ in our case. However, in some existing protocols~\cite{MTH17,TM17}, the number of settings is more than ${\rm poly}(n)$. b) The ratio of the number of randomly chosen registers in step 3 to that of the remaining registers must be less than $O(1/(n^4\log{n}))$. This is crucial in order for Theorems~\ref{soundness} and \ref{theorem2} to work. In an existing protocol~\cite{ZH18}, the number of the remaining (unmeasured) register is only one. It would be interesting to apply our analysis to these existing verification protocols~\cite{MTH17,TM17,ZH18} by appropriately modifying them. However, we leave this as an open problem for further research.

If the open problem will be solved, our analysis would become more attractive.
This is because hypergraph states are also resource states of MBQC~\cite{MM16,TMH18}.
Particularly, in Ref.~\cite{TMH18}, a hypergraph state $|G_n^d\rangle$ has been constructed such that it enables to perform universal MBQC with only Pauli-$X$ and $Z$ basis measurements. Note that the CV analogue of $|G_n^d\rangle$ can be efficiently verified using our protocol presented in this paper. This is because a CV hypergraph state is a special case of CV weighted hypergraph states.

Besides an application to the verification of universal quantum computing, our protocol can be applied to the verifiable blind quantum computing (VBQC) scenario. In VBQC, a client with weak quantum resources delegates an arbitrary quantum computing to a remote (universal) quantum server in such a way that the client's privacy is preserved and at the same time the integrity of the server is verified.
Almost all of the VBQC protocols can be divided into two types, i.e., remote state preparation (RSP) type~\cite{FK17} and measurement-only (MO) type~\cite{HM15}. In the former type, the client is required to prepare single- or multi-qubit quantum states. On the other hand, in the latter one, the client is required to perform single-qubit projective measurements on the quantum states (usually graph states). A non-verifiable blind quantum computing (BQC) protocol using CV graph states has already been proposed for both types~\cite{M12}, but a VBQC protocol using CV graph states has not yet been proposed. In this paper, we focus on the MO type because the homodyne measurement is generally considered to be significantly easier than the generation of highly squeezed states. By combining the original MO-type BQC protocol~\cite{MF13A} with our verification protocol of CV weighted hypergraph states, we construct a MO-type CV VBQC protocol as follows:
\begin{enumerate}
\item The quantum server generates $N_{\rm total}$ CV cluster states~\cite{MLGWRN06}, and sends them to the client.
\item The client performs our CV verification protocol. If the verification protocol succeeds (If Eq.~(\ref{threshold}) holds), they proceed to the next step. Otherwise, the client aborts the protocol.
\item The client performs MBQC on the verified quantum states.
\end{enumerate}
Note that this VBQC protocol requires no quantum memory for the client.
Although the protocol above is phrased sequentially, when no quantum memory is used, the computation is actually interspersed with the test runs which themselves take place out of order. The decision to accept or reject is taken once all qumodes have been measured or discarded.

From the universality of the CV cluster state, it follows that our CV VBQC protocol has perfect correctness, i.e., the client can obtain the correct result if the server is honest.
Next, our protocol is blind, i.e., the server cannot learn client's inputs, algorithms, and outputs, because there does not exist any communication channel from the client to the server.
Finally, the verifiability of our VBQC protocol follows from our CV verification protocol given in the RESULTS section.

In our CV VBQC protocol, we assume that the honest server sends the client an ideal CV graph state whose squeezing level is infinite. One possible approach to relax this assumption is to construct a test where even finitely squeezed states can pass with sufficiently high probability. Recently, Liu {\it et al.} have constructed a fidelity witness with respect to the tensor products of finitely squeezed cubic phase states~\cite{LDTAF18}. Based on this fidelity witness, they have also proposed another CV VBQC protocol. Therefore, their VBQC protocol does not need infinitely squeezed states unlike our VBQC protocol. On the other hand, their protocol assumes that the malicious server is restricted to preparing i.i.d. copies of single-qumode states while the malicious server in our protocol can perform any CPTP map as the attack. It would be interesting to combine these two approaches to design better and more practical CV VBQC protocols.\\

\medskip
\noindent{\bf\large METHODS}\\
In this section, we provide the primary mathematical tool used in the proof of Theorem~\ref{soundness} and an intuitive explanation of the proof. Our main tool is Serfling's bound:
\begin{lemma}[Serfling's bound~\cite{S74,TL17}]
\label{randomsampling}
Consider a set of binary random variables $Y=(Y_1,Y_2,\ldots,Y_T)$ with $Y_j$ taking values in $\{0,1\}$ and $T=N+K$. Then, for any $0<\nu<1$,
\begin{eqnarray}
{\rm Pr}\left[\sum_{j\in\bar{\Pi}}Y_j\ge \cfrac{N}{K}\sum_{j\in\Pi}Y_j+N\nu\right]\le{\rm exp}\left[-\cfrac{2\nu^2NK^2}{(N+K)(K+1)}\right],
\end{eqnarray}
where $\Pi$ is a set of $K$ samples chosen independently and uniformly at random from $Y$ without replacement. $\bar{\Pi}$ is the complementary set of $\Pi$.
\end{lemma}
Note that the sampling without replacement means that once a sample is selected, it is removed from the population in all subsequent selections.

In step 2 of our protocol, when $i=1$, Alice measures the stabilizer operator $g_1\in\{g_1^{(d)},g_1^{({\rm CV})}\}$ for $N_{\rm test}$ samples of the set $\Pi^{(1)}$ that is uniformly and randomly chosen out of the total $N_{\rm total}$ registers. In this case when Alice passes the stabilizer test for $g_1$ on the $j$th register $(1\le j\le N_{\rm test})$, $Y_j=0$; else $Y_j=1$. Therefore, by setting $K=N_{\rm test}$ and $T=N_{\rm total}$ in Lemma~\ref{randomsampling}, it reveals the upper bound
\begin{eqnarray}
\label{method1}
\cfrac{N_{\rm total}-N_{\rm test}}{N_{\rm test}}\sum_{j\in\Pi^{(1)}}Y_j+(N_{\rm total}-N_{\rm test})\nu
\end{eqnarray}
on the number of registers that are not stabilized by $g_1$ in the remaining complementary set $\bar{\Pi}^{(1)}$, which includes $N_{\rm total}-N_{\rm test}$ registers (for details, see the second paragraph of the proof in the Supplementary Information Section A).

Next, Alice performs the stabilizer tests for $g_2$ (when $i=2$) by uniformly and randomly choosing $N_{\rm test}$ registers, which are in the set $\Pi^{(2)}$, from $N_{\rm total}-N_{\rm test}$ remaining registers. 
Similarly, in Lemma~\ref{randomsampling}, by setting $K=N_{\rm total}$ and $T=N_{\rm total}-N_{\rm test}$, Alice can estimate that at most
\begin{eqnarray}
\label{method2}
\cfrac{N_{\rm total}-2N_{\rm test}}{N_{\rm test}}\sum_{j\in\Pi^{(2)}}Y_j+(N_{\rm total}-2N_{\rm test})\nu
\end{eqnarray}
registers are not stabilized by $g_2$ in the remaining $N_{\rm total}-2N_{\rm test}$ registers that are not measured by $g_1$ and $g_2$. 

From Eqs.~(\ref{method1}) and (\ref{method2}), we find a lower bound
\begin{eqnarray}
\nonumber
&&(N_{\rm total}-2N_{\rm test})-\left[\cfrac{N_{\rm total}-N_{\rm test}}{N_{\rm test}}\sum_{j\in\Pi^{(1)}}Y_j+(N_{\rm total}-N_{\rm test})\nu\right]\\
&&-\left[\cfrac{N_{\rm total}-2N_{\rm test}}{N_{\rm test}}\sum_{j\in\Pi^{(2)}}Y_j+(N_{\rm total}-2N_{\rm test})\nu\right]
\end{eqnarray}
on the number of remaining $N_{\rm total}-2N_{\rm test}$ registers that are stabilized by both $g_1$ and $g_2$. In estimating this lower bound, we use the pigeonhole principle. In other words, we consider a worst case scenario where the remaining registers that are not stabilized by $g_1$ and $g_2$ do not completely overlap with each other.

Using the same argument recursively, Alice estimates a lower bound on the number of the remaining registers stabilized by all of $\{g_i\}_{i=1}^n$ in the remaining $N_{\rm total}-nN_{\rm test}$ registers. Since only the ideal state $|G\rangle$ is stabilized by all the $g_i$ with $1\le i\le n$, this bound gives a lower bound $N_{\rm cor}^{\rm L}$ on the number of the ideal states in the remaining registers. If $N_{\rm cor}^{\rm L}/(N_{\rm total}-nN_{\rm test})$ is large, the averaged fidelity of the target register is also large because Alice finally selects one registers uniformly at random from the remaining registers.\\

\medskip
\noindent{\bf\large DATA AVAILABILITY}\\
No data sets were generated or analyzed during the current study.\\

\medskip
\noindent{\bf\large ACKNOWLEDGMENTS}\\
We thank Seiseki Akibue for helpful discussions.
Y.T. is supported by the Program for Leading Graduate Schools: Interactive Materials Science Cadet Program.
T.M. is supported by JST ACT-I No.JPMJPR16UP, the JSPS Grant-in-Aid for Young Scientists (B) No.JP17K12637, and JST PRESTO No.JPMJPR176A.
A. Mantri is supported by SUTD President's Graduate Fellowship. A. Mantri and
J.F.F. acknowledges support from Singapore's Ministry of Education and National Research Foundation, the ANR-NRF grant NRF2017-NRF-ANR004, and the US Air Force Office
of Scientific Research under AOARD grant FA2386-15-1-4082. This material is based on research funded in part by the Singapore National Research Foundation under NRF Award
NRF-NRFF2013-01.\\

\medskip
\noindent{\bf\large COMPETING INTERESTS}\\
J.F.F. has financial holdings in Horizon Quantum Computing Pte. Ltd.
{\red All other authors declare no competing interests.}\\

\medskip
\noindent{\bf\large CONTRIBUTIONS}\\
Y.T. conceived the key idea through discussions with A. Mantri and A. Mizutani. Y.T. refined the key idea through discussions with A. Mantri, T.M., and J.F.F. Y.T. performed the calculations, and all the authors contributed to checking the validity of the calculations and writing of the paper.\\

\medskip
\noindent{\red\bf\large CORRESPONDING AUTHOR}\\
{\red Correspondence to Yuki Takeuchi.}

\newpage
\noindent
{\bf\large Supplementary Information for Resource-efficient verification of quantum computing using Serfling's bound}\\
In Section~\ref{A}, we give a proof of Theorem~1.
In Section~\ref{B}, we give a proof of Theorem~2.

\subsection{Proof of Theorem~1}
\label{A}
In this section, we give a proof of Theorem~1. \\
{\it Proof.}
From Lemma~1, after Alice performs the stabilizer tests for $g_i\in\{g_i^{(d)},g_i^{({\rm CV})}\}$ on all the $N_{\rm test}$ registers, we have
\begin{eqnarray}
\label{pr1}
&&{\rm Pr}\left[\sum_{j\in\bar{\Pi}^{(i)}}Y_j<(N_{\rm total}-iN_{\rm test}) \left(\cfrac{\sum_{j\in\Pi^{(i)}}Y_j}{N_{\rm test}}+\nu\right)\right]\ \ \ \ \ \\
\nonumber
&>&1-{\rm exp}\left\{-\cfrac{2\nu^2(N_{\rm total}-iN_{\rm test})N_{\rm test}^2}{[N_{\rm total}-(i-1)N_{\rm test}](N_{\rm test}+1)}\right\}\\
\nonumber
&=&1-{\rm exp}\left[-\cfrac{2\nu^2N_{\rm test}}{1+N_{\rm test}/(N_{\rm total}-iN_{\rm test})}\cfrac{1}{1+1/N_{\rm test}}\right]\\
\nonumber
&=&1-{\rm exp}\left[-2\nu^2N_{\rm test}\cfrac{1}{1+1/(2n-i)}\cfrac{1}{1+1/N_{\rm test}}\right]\\
\label{pr2}
&\equiv&q_i,
\end{eqnarray}
where $\Pi^{(i)}$ is a set of registers used to perform the stabilizer test for $g_i$, $\bar{\Pi}^{(i)}$ is a set of the remaining registers after finishing the stabilizer tests for $g_i$, and we have used $N_{\rm total}=2nN_{\rm test}$ to derive the last equality. Let us set $\nu=\sqrt{c}/n^2$, where $c$ is any constant satisfying $64/5<c<(n-1)^2/4$. Let $N_{\rm pass}\equiv\sum_{i=1}^nN_{{\rm pass},i}$. We can guarantee, from Eqs.~(\ref{pr1}) and (\ref{pr2}), that at least
\begin{eqnarray}
\label{number}
&&(N_{\rm total}-nN_{\rm test})-\sum_{i=1}^n(N_{\rm total}-iN_{\rm test})\left(\cfrac{\sum_{j\in\Pi^{(i)}}Y_j}{N_{\rm test}}+\nu\right)\\
\nonumber
&\ge&nN_{\rm test}-\nu nN_{\rm total}-\cfrac{N_{\rm total}}{N_{\rm test}}\sum_{i=1}^n\sum_{j\in\Pi^{(i)}}Y_j\\
\nonumber
&=&nN_{\rm test}-\nu nN_{\rm total}-\cfrac{N_{\rm total}}{N_{\rm test}}(nN_{\rm test}-N_{\rm pass})\\
\nonumber
&=&\left(n-2\nu n^2-2n^2+\cfrac{2nN_{\rm pass}}{N_{\rm test}}\right)N_{\rm test}\\
\label{remainreg}
&=&\left(n-2\sqrt{c}-2n^2+\cfrac{2nN_{\rm pass}}{N_{\rm test}}\right)N_{\rm test}
\end{eqnarray}
registers would pass all of the stabilizer tests (if they were performed) with probability larger than
\begin{eqnarray}
\nonumber
\prod_{i=1}^nq_i&\ge&q_n^n\\
\nonumber
&=&\left\{1-{\rm exp}\left[-2\nu^2N_{\rm test}\cfrac{1}{1+1/n}\cfrac{1}{1+1/N_{\rm test}}\right]\right\}^n\\
\nonumber
&\ge&\left[1-{\rm exp}\left(-\cfrac{\nu^2N_{\rm test}}{2}\right)\right]^n\\
\nonumber
&\ge&\left[1-{\rm exp}\left(-\cfrac{5c}{64}\log{n}\right)\right]^n\\
\nonumber
&=&\left(1-n^{-5c/64}\right)^n\\
\label{probability}
&>&1-n^{1-5c/64},
\end{eqnarray}
where we have used $n\ge 1$ and $N_{\rm test}\ge 1$ to derive the second inequality. Note that we assume $n\ge 9$ in Theorem~1, but in order to simplify the calculation, we here use $n\ge 1$. 

Let us apply Eqs.~(\ref{remainreg}) and (\ref{probability}) to an $n$-qubit graph state $|G\rangle$. It is known that 
\begin{eqnarray}
\label{basis}
\left\{|G({\bf a})\rangle\equiv\prod_{i=1}^n{Z_i}^{a_i}|G\rangle\Bigg|\forall i, a_i\in\{0,1\}\right\}
\end{eqnarray}
is the orthonormal basis, where $Z_i$ is the Pauli-$Z$ operator acting on the $i$th qubit, and ${\bf a}\equiv\{a_i\}_{i=1}^n$. It means that any quantum state can be expanded by these basis states. Eqs.~(\ref{remainreg}) and (\ref{probability}) means that when the state $\rho$ of the remaining $N_{\rm total}-nN_{\rm test}(\equiv N_{\rm rest})$ registers is expanded by these basis states:
\begin{eqnarray*}
\rho=\sum_{\vec{{\bf a}},\vec{{\bf a}}'}p_{\vec{{\bf a}},\vec{{\bf a}}'}\left(\bigotimes_{k=1}^{N_{\rm rest}}|G({\bf a}_k)\rangle\right)\left(\bigotimes_{k=1}^{N_{\rm rest}}\langle G({\bf a'}_k)|\right),
\end{eqnarray*}
where $\vec{{\bf a}}\equiv({\bf a}_1,\ldots,{\bf a}_{N_{\rm rest}})$ and $\sum_{\vec{{\bf a}}}p_{\vec{{\bf a}},\vec{{\bf a}}}=1$,
the sum $\sum_{\vec{{\bf a}}\in S}p_{\vec{{\bf a}},\vec{{\bf a}}}$ is larger than $1-n^{1-5c/64}$. Here, $S$ is a set of $\vec{{\bf a}}$ such that the number of ${\bf a}_k$ all of whose entries are $0$ is larger than $(n-2\sqrt{c}-2n^2+2nN_{\rm pass}/N_{\rm test})N_{\rm test}$.

With respect to qudit graph states, we can apply the same argument by replacing Eq.~(\ref{basis}) with the qudit orthonormal basis
\begin{eqnarray*}
\left\{\prod_{i=1}^n{Z_i}^{a_i}|G_d\rangle\Bigg|\forall i, a_i\in\{0,1,\ldots,d-1\}\right\},
\end{eqnarray*}
where $Z_i$ is the qudit Pauli-$Z$ operator acting on the $i$th qudit. With respect to CV weighted hypergraph states, we can also apply the same argument by replacing Eq.~(\ref{basis}) with the CV orthonormal basis
\begin{eqnarray*}
\left\{\prod_{i=1}^n{Z_i(s_i)}|G_{\rm CV}\rangle\Bigg|\forall i, s_i\in\mathbb{R}\right\},
\end{eqnarray*}
where $Z_i(s_i)\equiv e^{is_i\hat{x}_i}$ is the Weyl-Heisenberg operator acting on the $i$th qumode.

Since only the ideal state $|G\rangle(\in\{|G_d\rangle,|G_{\rm CV}\rangle\})$ always passes the stabilizer test for $g_i$ for any $i$, the ratio of the number of ideal states to that of non-ideal quantum states in the remaining registers is larger than
\begin{eqnarray*}\cfrac{(n-2\sqrt{c}-2n^2+2nN_{\rm pass}/N_{\rm test})N_{\rm test}}{N_{\rm total}-nN_{\rm test}}=1-\cfrac{2\sqrt{c}}{n}-2n\left(1-\cfrac{N_{\rm pass}}{nN_{\rm test}}\right).
\end{eqnarray*}
Since the uniform random selection in step 3 is equivalent to selecting the first register of the remaining registers after the random permutation, the averaged state $\rho_{\rm tgt}$ of the target register (over all random permutations) is written as
\begin{eqnarray}
\label{tgt}
\rho_{\rm tgt}=\left[1-\cfrac{2\sqrt{c}}{n}-2n\left(1-\cfrac{N_{\rm pass}}{nN_{\rm test}}\right)\right]|G\rangle\langle G|+\ldots
\end{eqnarray}
in the worst case where Eq.~(\ref{number}) is minimized. Note that off diagonal elements do not affect the fidelity with $|G\rangle$ because $|G\rangle$ is orthogonal to other orthonormal basis states. From Eqs.~(\ref{probability}) and (\ref{tgt}), we finally conclude that
\begin{eqnarray*}
\langle G|\rho_{\rm tgt}|G\rangle\ge1-\cfrac{2\sqrt{c}}{n}-2n\left(1-\cfrac{N_{\rm pass}}{nN_{\rm test}}\right)
\end{eqnarray*}
with probability larger than $1-n^{1-5c/64}$.
\hspace{\fill}$\blacksquare$

\medskip
\subsection{Proof of Theorem~2}
\label{B}
In this section, we give a proof of Theorem~2.\\
{\it Proof.}
We modify our protocol in Theorem~1 in such a way that Alice chooses $\tilde{n}$ registers uniformly at random in step 3. Let $N_{\rm rest}$ and $N_{\rm cor}^{\rm L}$ be the number of remaining registers after step 2 and the lower bound on the number of the correct states in the remaining registers, respectively. If we set parameters $N_{\rm total}$ and $N_{\rm test}$ as with Theorem~1, from Eqs.~(\ref{remainreg}) and (\ref{probability}),
\begin{eqnarray*}
N_{\rm rest}&=&nN_{\rm test},\\
N_{\rm cor}^{\rm L}&=&{\rm max}\left\{\left\lceil\left(n-2\sqrt{c}-2n^2+2nN_{\rm pass}/N_{\rm test}\right)N_{\rm test}\right\rceil,0\right\}
\end{eqnarray*}
are satisfied with probability at least $1-n^{1-5c/64}$.
In this case, the averaged state $\tilde{\rho}_{\rm tgt}$ of the $\tilde{n}$ registers can be written as
\begin{eqnarray*}
\tilde{\rho}_{\rm tgt}=\prod_{i=0}^{\tilde{n}-1}\cfrac{N_{\rm cor}^{\rm L}-i}{N_{\rm rest}-i}|G\rangle\langle G|^{\otimes \tilde{n}}+\ldots.
\end{eqnarray*}
Therefore, we obtain
\begin{eqnarray}
\nonumber
\langle G|^{\otimes \tilde{n}}\tilde{\rho}_{\rm tgt}|G\rangle^{\otimes \tilde{n}}&\ge&\prod_{i=0}^{\tilde{n}-1}\cfrac{N_{\rm cor}^{\rm L}-i}{N_{\rm rest}-i}\\
\label{fidelity_multi}
&\ge&\left(\cfrac{N_{\rm cor}^{\rm L}-\tilde{n}+1}{N_{\rm rest}-\tilde{n}+1}\right)^{\tilde{n}}.
\end{eqnarray}
If $\tilde{n}=O(n^t)$, where $t<1$,
\begin{eqnarray}
\nonumber
&&\left(\cfrac{N_{\rm cor}^{\rm L}-\tilde{n}+1}{N_{\rm rest}-\tilde{n}+1}\right)^{\tilde{n}}\\
\nonumber
&=&\left(1-\cfrac{N_{\rm rest}-N_{\rm cor}^{\rm L}}{N_{\rm rest}-\tilde{n}+1}\right)^{\tilde{n}}\\
\label{fidelity_multi2}
&\ge&1-\cfrac{(2\sqrt{c}+2n^2-2nN_{\rm pass}/N_{\rm test})\tilde{n}{\red N_{\rm test}}}{\red nN_{\rm test}-(\tilde{n}-1)}.
\end{eqnarray}
By combining Eqs.~(\ref{fidelity_multi}) and (\ref{fidelity_multi2}), Theorem~2 is derived.
\hspace{\fill}$\blacksquare$

\end{document}